\documentclass{article}
\usepackage[T1]{fontenc}
\usepackage[latin9]{inputenc}
\usepackage{color}
\usepackage{tipa}
\usepackage{amssymb}
\usepackage{graphicx}
\usepackage{setspace}
\usepackage{esint}
\begin{document}

\title{\textcolor{black}{Inertia from an Asymmetric Casimir Effect.}}

\author{\textcolor{black}{M.E. McCulloch }%
\thanks{\textcolor{black}{SMSE, Plymouth University, Plymouth, PL4 8AA, UK.
mike.mcculloch@plymouth.ac.uk}%
}}
\maketitle
\begin{abstract}
\textcolor{black}{The property of inertia has never been fully explained.
A model for inertia (MiHsC or quantised inertia) has been suggested
that assumes that 1) inertia is due to Unruh radiation and 2) this
radiation is subject to a Hubble-scale Casimir effect. This model
has no adjustable parameters and predicts the cosmic acceleration,
and galaxy rotation without dark matter, suggesting that Unruh radiation
indeed causes inertia, but the exact mechanism by which it does this
has not been specified. The mechanism suggested here is that when
an object accelerates, for example to the right, a dynamical (Rindler)
event horizon forms to its left, reducing the Unruh radiation on that
side by a Rindler-scale Casimir effect whereas the radiation on the
other side is only slightly reduced by a Hubble-scale Casimir effect.
This produces an imbalance in the radiation pressure on the object,
and a net force that always opposes acceleration, like inertia. A
formula for inertia is derived, and an experimental test is suggested.}
\end{abstract}

\section{\textcolor{black}{Introduction}}

\textcolor{black}{Hawking (1975) showed that the event horizon of
a black hole can separate paired virtual particles leading to Hawking
radiation. It was also proposed by Fulling (1973), Davies (1975) and
Unruh (1976) that a similar effect happens to accelerated objects
in that a dynamical Rindler event horizon forms on the side they are
accelerating away from. This horizon similarly produces radiation
so that an accelerated object will perceive a warm background full
of blackbody radiation whereas a non-accelerated observer in the same
space will see a cold background with no radiation. The Unruh temperature
seen by a body with an acceleration 'a' is given by $T=\hbar a/2\pi ck$,
where \textcrh{} is the reduced Planck's constant, c is the speed
of light and k is Boltzmann's constant. This is now called the Fulling-Davies-Unruh
effect, or the Unruh effect for short, and it is unclear whether it
has been observed or not. It may be the explanation for the observed
Sokolov-Turnov effect (Akhmedov and Singleton, 2007).}

\textcolor{black}{The property known as inertia has never been adequately
explained, and has been rather a neglected part of physics. One model
for inertia that uses the electromagnetic part of the Unruh radiation
was suggested by Haisch }\textit{\textcolor{black}{et al}}\textcolor{black}{.
(1994). They proposed that oscillating partons within an accelerated
object feel a magnetic Lorentz force, due to their interaction with
the zero-point field, that opposes the acceleration of the object.
The force they derived from this model was $F=-\Gamma w_{c}^{2}\hbar a/2\pi c^{2}$
where $\Gamma$ is the Abraham-Lorentz damping constant of the parton
being oscillated, $w_{c}$ is the Compton scale of the parton below
which the oscillations of the zero-point field have no effect on it,$\hbar$
is the reduced Planck's constant, a is the acceleration and c is the
speed of light. However, although their derived force looks like inertia,
their derivation was complex, required the imposition of a high frequency
cutoff to avoid infinite energy, and has been criticised on relativistic,
and other, grounds, by, for example, Levin (2009).}

\begin{singlespace}
\noindent \textcolor{black}{McCulloch (2007) proposed a model for
inertia that could be called a Modification of inertia resulting from
a Hubble-scale Casimir effect (MiHsC) or Quantised Inertia. MiHsC
assumes that the inertial mass of an object is caused (somehow) by
Unruh radiation resulting from its acceleration with respect to surrounding
matter, and that this radiation is subject to a Hubble-scale Casimir
effect. This means that only Unruh waves that fit exactly into twice
the Hubble diameter are allowed, so that an increasingly greater proportion
of the Unruh waves are disallowed as they get longer (as acceleration
decreases), leading to a new gradual loss of inertia as acceleration
reduces. In MiHsC the inertial mass becomes}

\noindent \textcolor{black}{
\begin{equation}
m_{I}=m_{g}\left(1-\frac{\beta\pi^{2}c^{2}}{|a|\Theta}\right)\sim m_{g}\left(1-\frac{2c^{2}}{|a|\Theta}\right)
\end{equation}
}

\noindent \textcolor{black}{where $m_{g}$ is the gravitational mass,
$\beta=0.2$ (from Wien's displacement law), c is the speed of light,
and $\Theta$ is the Hubble diameter ($2.7\times10^{26}m$, from Freedman
}\textit{\textcolor{black}{et al}}\textcolor{black}{., 2001). For
the derivation of Eq. 1 see McCulloch (2007) and for a justification
for the use of the modulus of the acceleration see McCulloch (2008b).
MiHsC has no adjustable parameters, and predicts cosmic acceleration
and galaxy rotation without dark matter (see McCulloch, 2007, 2010,
2012). It violates the equivalance principle, but not in a way that
could have been detected in torsion balance experiments (McCulloch,
2011). It has been suggested by Gine (2012) that there may be a link
between MiHsC and holographic entropic gravity models (eg: both use
the Unruh temperature), but this possible link is not yet clear.}
\end{singlespace}

\textcolor{black}{The agreements between MiHsC and anomalous observations
provide support to suggest that inertia is (somehow) proportional
to the energy in the Unruh radiation spectrum, but an exact mechanism
has not been proposed. In this paper for the first time a specific
mechanism is suggested.}

\section{\textcolor{black}{Method \& Results}}

\textcolor{black}{When an object (O) is accelerated to the right as
shown in the schematic (Fig. 1), Unruh radiation appears anisotropically
and hits the object from all directions, but a dynamic (Rindler) event
horizon forms on the left side, since information from the region
of space behind this event horizon can never hope to catch up with
the object (see the shaded area in the schematic). Now if we calculate
the energy density of the Unruh radiation, in the direction of acceleration,
to the right, most of the Unruh waves will be allowed by the Hubble-scale
Casimir effect (McCulloch, 2007) since the event horizon is far away
at the Hubble distance, but on the opposite side, to the left, fewer
waves in the Unruh spectrum will be allowed because the dynamic (Rindler)
event horizon is much closer, at a distance of $c^{2}/a$ (where a
is the acceleration, see eg: Rindler, 2001) so the momentum impact
of the Unruh radiation will be lower from the left (with a Rindler-scale
Casimir effect) and greater from the right (with a Hubble-scale Casimir
effect) and this will push the object back against the applied acceleration.
This asymetric Casimir effect models inertia intuitively.}

\textcolor{black}{The specific calculation can be done as follows.
A single particle (O) is considered for simplicity. It is accelerating
to the right as shown in the schematic (Fig. 1). The radiation pressure
(force) on any small surface area (A) exposed to anisotropic radiation,
like Unruh radiation, is}

\textcolor{black}{
\begin{equation}
F=\frac{uA}{3}
\end{equation}
}

\textcolor{black}{where u is the radiation energy density, and A is
the surface area intercepting this radiation, which is assumed to
be a small part of the surface area of the whole particle. Now we
need to look at the net difference between the force from the left
and the right, so to start with, we can consider a line through the
particle at an arbitrary angle $\theta$(see the long dashed line
in Fig. 1) and calculate the net force onto the particle along this
line from both directions (see arrows) and take its component along
the x-axis}

\textcolor{black}{
\begin{equation}
dF_{x}=\frac{u_{left}Acos\theta}{3}-\frac{u_{right}Acos\theta}{3}
\end{equation}
}

\textcolor{black}{The energy in the Unruh radiation coming from the
right (the second term on the right hand side) is subject to the usual
Hubble-scale Casimir effect of McCulloch (2007), so $u_{right}'=u(1-\lambda/4\Theta))$,
where $\lambda$ is the peak wavelength of the Unruh spectrum. In
contrast, the energy of the radiation coming from the left (the first
term in the right hand side) is subject to a Rindler-scale Casimir
effect with the event horizon now at a much smaller distance away
of $c^{2}/acos\theta$ (where $acos\theta$ is the component of the
acceleration in the direction $\theta$) so that $u_{left}'=u(1-\lambda/4(c^{2}/acos\theta))$.
The difference, the net force in the x-direction, is}

\textcolor{black}{
\begin{equation}
dF_{x}=\frac{uAcos\theta}{3}\left(1-\frac{\lambda acos\theta}{4c^{2}}-1+\frac{\lambda}{4\Theta}\right)
\end{equation}
}

\textcolor{black}{This simplifies to}

\textcolor{black}{
\begin{equation}
dF_{x}=\frac{u\lambda Acos\theta}{3}\left(\frac{1}{4\Theta}-\frac{acos\theta}{4c^{2}}\right)
\end{equation}
}

\textcolor{black}{We now integrate the contribution from all possible
angles. To do this we integrate from $\theta=0$ to $\theta=\pi/2$,
then double this to get the result for the x-y plane and then integrate
this circularily $180^{o}$ or $\pi$ around the x axis (around the
angle $\phi$) to calculate the total force}

\textcolor{black}{
\begin{equation}
dF_{x}=2\times\frac{u\lambda A}{3}\int_{0}^{\pi}\int_{0}^{\pi/2}\left(\frac{cos\theta}{4\Theta}-\frac{acos^{2}\theta}{4c^{2}}\right)d\theta d\phi
\end{equation}
}

\textcolor{black}{
\begin{equation}
dF_{x}=\frac{2u\lambda A}{3}\int_{0}^{\pi}\left[\frac{sin\theta}{4\Theta}-\frac{a\theta}{8c^{2}}-\frac{asin2\theta}{16c^{2}}\right]_{0}^{\pi/2}d\phi
\end{equation}
}

\textcolor{black}{The circular integral over $\phi$ is equivalent
to a multiplication by $\pi$}

\textcolor{black}{
\begin{equation}
dF_{x}=\pi\times\frac{2u\lambda A}{3}\left[\frac{1}{4\Theta}-\frac{\pi a}{16c^{2}}\right]
\end{equation}
}

\textcolor{black}{The first term is the brackets is the MiHsC correction
to inertia (see McCulloch, 2007) and is tiny compared to the second
term, except for low accelerations. Assuming a large acceleration,
ie: a terrestrial one, we can neglect this MiHsC term}

\textcolor{black}{
\begin{equation}
dF_{x}=-\frac{\pi^{2}u\lambda Aa}{24c^{2}}
\end{equation}
}

\textcolor{black}{Since $u=E/V=hc/\lambda V$ where V is volume, then}

\textcolor{black}{
\begin{equation}
dF_{x}=-\frac{\pi^{2}hAa}{24cV}
\end{equation}
}

\textcolor{black}{This is taken to be the force on one spherical particle.
It implies that the Rindler event horizon that forms in the reference
frame of an accelerated particle reduces the energy density of the
Unruh radiation in the direction opposite to the acceleration vector,
so that it is unable to balance the momentum transferred by the Unruh
radiation from the other direction. This produces a net force that
is always counter to the acceleration (the minus sign in the above
formula), and this is a characteristic of inertia. }

\textcolor{black}{The ratio A/V could be simplified by choosing a
distance scale (x) smaller than the particle so that $A/V=x^{2}/x^{3}=1/x$.
Using the Planck distance ($l_{P}=1.616\times10^{-35}m$) as x, for
example, then Eq. 10 becomes}

\textcolor{black}{
\begin{equation}
dF_{x}=-\frac{\pi^{2}ha}{24cl_{P}}
\end{equation}
}

\textcolor{black}{and the inertial mass is $m_{i}\sim\pi^{2}h/24cl_{P}\sim5.5\times10^{-8}kg$
which is about twice the Planck mass ($m_{P}=2.176\times10^{-8}kg$).
This derivation is only valid for single particles. To calculate the
inertial mass of a compound object it would be necessary to multiply
the above mass by the number of particles present. It is also only
valid for velocities much slower than that of light, since otherwise
Lorentz contraction would alter the dimensions of the particle.}

\section{\textcolor{black}{Discussion}}

\textcolor{black}{The formula for inertia derived here (Eq. 11) is
considerably simpler, in both derivation and final form, than the
formula of Haisch }\textit{\textcolor{black}{et al}}\textcolor{black}{.
(1994) and the mechanism is different. For Haisch et al. (1994) the
inertial process was a magnetic Lorentz force acting on particles
oscillating at very high frequencies, and to avoid an infinite energy
they had to impose an upper frequency limit. The asymetric Casimir
effect proposed here produces an energy difference, so no cutoff is
needed.}

\textcolor{black}{This explanation for inertia depends upon the existence
of Unruh radiation, which has not been directly observed. However,
the results in this paper, and previous works (McCulloch, 2007-2012)
have shown that if one is willing to accept the existence of Unruh
radiation, apply Rindler- and Hubble-scale Casimir effects to it,
and allow it to have an impact on objects, then certain observed anomalies
can be explained simply, and, as shown here, a model for inertia can
be derived. These results provide some indirect evidence for Unruh
radiation.}

\textcolor{black}{The implied dependence of inertia on event horizons
suggests a way to test this idea. It was suggested by McCulloch (2008),
that it may be possible to modify inertia by creating an event horizon
using the metamaterials that were proposed by Pendry et al. (2006)
and Leonhardt (2006). They have demonstrated theoretically that radiation
of a given wavelength can be bent around an object (which must be
smaller than the wavelength) using a metamaterial, making that object
invisible to an observer at that wavelength. It may be possible instead,
to set up a metamaterial to reflect radiation in such a way that an
artificial event horizon is formed. Then according to the model discussed
here, this will damp Unruh radiation on one side of the object which
would then be accelerated towards the metamaterial. It may be simpler
to use metamaterials to more directly damp the Unruh radiation on
one side of the object.}

\section{\textcolor{black}{Conclusions}}

\textcolor{black}{A new model for inertia (MiHsC or quantised inertia)
has been suggested that assumes that 1) inertia is due to Unruh radiation
and 2) this radiation is subject to a Hubble-scale Casimir effect.
Here, for the first time, a mechanistic model for MiHsC, and inertia,
is suggested.}

\textcolor{black}{The model assumes that when an object accelerates
in one direction, a dynamical Rindler event horizon forms in the opposite
direction, producing a Casimir effect, that reduces the Unruh radiation
there. As a result, the Unruh radiation pressure on the object is
greater from the direction of acceleration, producing a net force
that always opposes acceleration, just like inertia.}

\textcolor{black}{This model for inertia suggests that if some way
could be found to damp the Unruh waves on one side of an object, or
create an artificial event horizon on that side (perhaps using metamaterials),
the object could then be accelerated in a new way.}

\section*{\textcolor{black}{Acknowledgements}}

\textcolor{black}{Many thanks to T. O'Hare for useful conversations,
to the anonymous reviewers for their comments, and to B. Kim for support
and encouragement.}

\section*{\textcolor{black}{References}}

\textcolor{black}{Akhmedov, E.T., and D. Singleton, 2007. On the physical
meaning of the Unruh effect. }\textit{\textcolor{black}{JETP Letters}}\textcolor{black}{,
86(9), 615-619.}

\textcolor{black}{Davies, P.C.W., 1975. Scalar production in Schwartzschild
and Rindler metrics. }\textit{\textcolor{black}{J. Phys. A}}\textcolor{black}{,
8(4), 609.}

\textcolor{black}{Freedman, W.L. }\textit{\textcolor{black}{et al}}\textcolor{black}{.,
2001. Final results from the Hubble Space Telescope key project to
measure the Hubble constant.}\textit{\textcolor{black}{{} ApJ}}\textcolor{black}{,
553, 47-72.}

\textcolor{black}{Fulling, S.A., 1973. Nonuniqueness of canonical
field quantization in Riemannian space-time. }\textit{\textcolor{black}{Phys.
Rev. D}}\textcolor{black}{., 7(10), 2850.}

\textcolor{black}{Gine, J., 2012. The holographic scenario, the modified
inertia and the dynamics of the universe. }\textit{\textcolor{black}{Mod.
Phys. Lett. A}}\textcolor{black}{. Vol. 27, No. 34, 1250208.}

\textcolor{black}{Haisch, B., A. Rueda and H.E. Puthoff, 1994. Inertia
as a zero-point field Lorentz force. }\textit{\textcolor{black}{Phys.
Rev. A}}\textcolor{black}{., 49, 678.}

\textcolor{black}{Hawking, S., 1974. Black hole explosions. }\textit{\textcolor{black}{Nature}}\textcolor{black}{,
248, 30.}

\textcolor{black}{Leonhardt, U., 2006. Optical conformal mapping.
}\textit{\textcolor{black}{Science}}\textcolor{black}{, 312, 1777-1780.}

\textcolor{black}{Levin, Y.S., 2009. Inertia as a zero-point field
force: critical analysis of the Haisch, Rueda, Puthoff inertia theory.
}\textit{\textcolor{black}{Phys. Rev. A}}\textcolor{black}{, 79(1),
012114.}

\textcolor{black}{McCulloch, M.E., 2007. The Pioneer anomaly from
modified inertia. }\textit{\textcolor{black}{Mon. Not Roy. Astro.
Soc.}}\textcolor{black}{, 376, 338.}

\textcolor{black}{McCulloch, M.E., 2008a. Can the flyby anomalies
be explained using a modification of inertia? }\textit{\textcolor{black}{JBIS}}\textcolor{black}{,
61, 373-378.}

\begin{singlespace}
\noindent \textcolor{black}{McCulloch, M.E., 2008b. Modelling the
flyby anomalies using a modification of inertia. }\textit{\textcolor{black}{MNRAS-letters}}\textcolor{black}{,
389(1), L57-60.}

\noindent \textcolor{black}{McCulloch, M.E., 2010. Minimum accelerations
from quantised inertia. }\textit{\textcolor{black}{EPL}}\textcolor{black}{,
90, 29001.}
\end{singlespace}

\textcolor{black}{McCulloch, M.E., 2011. The Tajmar effect from quantised
inertia. }\textit{\textcolor{black}{EPL}}\textcolor{black}{, 95, 39002.}

\textcolor{black}{McCulloch, M.E., 2012. Testing quantised inertia
on galactic scales. }\textit{\textcolor{black}{Astrophysics and Space
Science}}\textcolor{black}{, 342, 2, 575-578.}

\textcolor{black}{Pendry, J.B., D. Schurig, and D.R. Smith, 2006.
Controlling electromagnetic fields. }\textit{\textcolor{black}{Science}}\textcolor{black}{,
312, 1780-1782.}

\textcolor{black}{Rindler, W., 2001. Relativity, special, general
and cosmological. Oxford University Press.}

\textcolor{black}{Unruh, W.G., 1976. Notes on black hole evaporation.
}\textit{\textcolor{black}{Phys. Rev. D}}\textcolor{black}{., 14,
870-892.}

\section*{\textcolor{black}{Figures}}

\textcolor{black}{\includegraphics[scale=0.8]{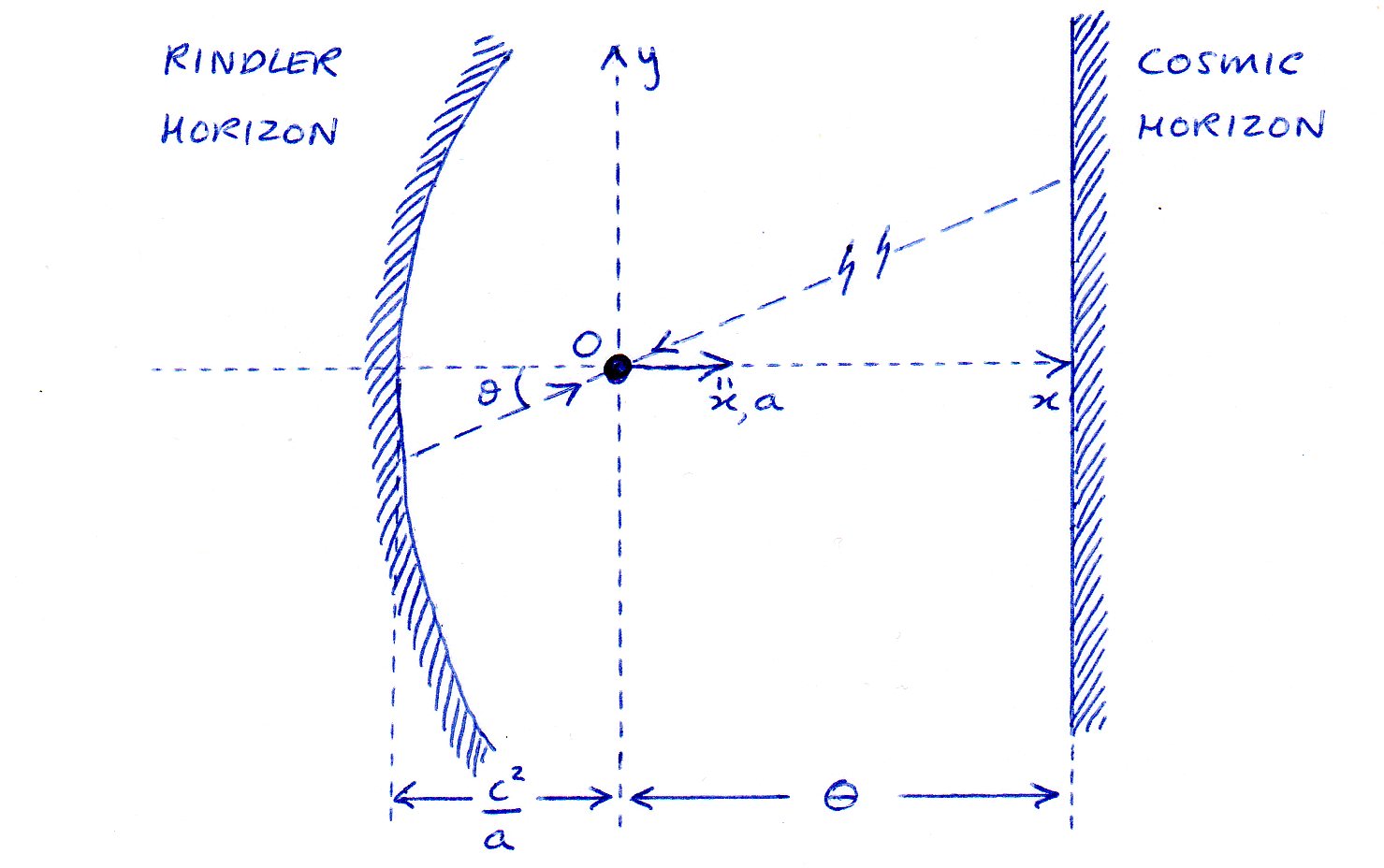}}

\textcolor{black}{Figure 1. A schematic showing a particle accelerating
rightwards (O). The shading shows the cosmic horizon far away to its
right (at a distance $\Theta/2$) and a closer Rindler horizon to
its left (at a distance $c^{2}/a$ away). This produces an asymetric
Casimir effect that pushes the particle (O) to the left against its
acceleration: a model for inertia. The lower case $\theta$ shows
the angle of integration in the x,y plane.}
\end{document}